**Ultra High Thermal Resolution Scanning Probe Microscopy via Carbon Nanotube Tipped Thermal Probes**

*Peter D. Tovee[a†], Manuel E. Pumarol[a†], Mark C. Rosamond[b†], Robert Jones[a], Michael C. Petty[b], Dagou A. Zeze[b] and Oleg V. Kolosov*[a]*

*[a]Physics Department, Lancaster University, Lancaster, LA1 4YB, UK. E-mail: o.kolosov@lancaster.ac.uk*
*[b]School of Engineering & Computer Sciences, Durham University, Durham, DH1 3LE, UK*
*[†]These authors contributed equally to the paper.*

We present a new concept of scanning thermal nanoprobe that utilizes the extreme thermal conductance of a carbon nanotube (CNT) to channel heat between the probe and the sample. The integration of CNT in scanning thermal microscopy (SThM) overcomes the main drawbacks of standard SThM probes, where the low thermal conductance of the apex SThM probe is the main limiting factor. The integration of CNT (CNT- SThM) extends SThM sensitivity to thermal transport measurement in higher thermal conductivity materials such as metals, semiconductors and ceramics, while also improving the spatial resolution. Investigation of thermal transport in ultra large scale integration (ULSI) interconnects, using CNT- SThM probe, showed fine details of heat transport in ceramic layer, vital for mitigating electromigration in ULSI metallic current leads. For a few layer graphene, the heat transport sensitivity and spatial resolution of the CNT-SThM probe demonstrated significantly superior thermal resolution compared to that of standard SThM probes achieving 20-30 nm topography and ~30 nm thermal spatial resolution compared to 50-100 nm for standard SThM probes. The outstanding axial thermal conductivity, high aspect ratio and robustness of CNTs can make CNT-SThM the perfect thermal probe for the measurement of nanoscale thermophysical properties and an excellent candidate for the next generation of thermal microscopes.

## I. Introduction

The constantly developing variety of scanning probe microscopy (SPM)[1, 2] techniques enables the investigation of important features in the nanoworld, from the nanoscale resolution topographical mapping of semiconductors and protein macromolecules,[3] to exploration of nanoscale electronic[4, 5], mechanical,[6-9] magnetic,[10] and thermal properties.[11, 12] Scanning thermal microscopy (SThM)[11-19] has come a long way since its invention[11] in 1986 and currently plays a leading role in the investigation of thermal properties on the nanoscale. SThM use a self-heating thermal sensors incorporated within a sharp tip that is thermally contacted with a sample surface and is widely used in studies of polymeric and organic materials[15]. While SThM also allows the study of the temperature distribution and heat transport in nanodevices such as interconnects, memory and RF circuits, phase change memory and semiconductor laser structures, high thermal conductivities of semiconductors and metals constitutes a major challenge for both spatial resolution and sensitivity of SThM with the main limiting factor being the low efficiency of the thermal coupling between the temperature sensing element and the local area of the studied surface.

The lateral resolution of SThM has never reached that of mainstream atomic force microscopy (AFM) methods, mainly due to the more bulky structure of the sensor, which incorporates either the resistive heating element, such as in the Wollaston wire probes[20] (WW), silicon nitride with resistive Pd layer[21] (SP) probe, micromanufactured doped Si[22, 23] (DS) probe, or a thermocouple.[24] The heat dissipation by the liquid layer,[25] to ambient air, and via radiation can further deteriorate the resolution. At the same time, it has been shown elsewhere that SThM sensitivity to thermal transport of higher thermal conductivity materials is mainly limited by the thermal resistance of the tip apex,[14, 26] and can be only partially improved by reducing heat leaks to the environment, e.g. via vacuum enclosure.[18, 19, 27] In this paper, we exploit the unique properties of carbon nanotubes (CNTs), which are well known for their superior mechanical properties, high aspect ratio[18, 28] and very high axial thermal conductivity,[28] to create a superior SThM sensor capable of high performance thermal transport measurements. While CNT probes were used in SPM[29-31] resulting in commercial CNT probes for standard contact AFM,[32] and CNTs have been used for local heating in data storage experiments,[33] to the best of our knowledge there are no reports to date on heat transport measurements in SThM imaging using such probes. Here, we describe the pathway to assemble CNT-SThM probes for use in thermal sensing and imaging, and use their superior thermal sensitivity to explore heat transport of complex metallic-ceramic-polymeric nanostructures of ultra large scale integration (ULSI) interconnects.[34] We then investigate a heat transport of highly thermally conductive few-layer-graphene (FLG)[35] on lower, but still highly thermally conductive Si substrate, achieving a differentiation between these materials and the thermal spatial resolution of a few tens of nm in an ambient environment, with superior topographical and friction resolution compared to the same probe without the CNT tip.

## II. Materials and methods

### 1. CNT-SThM probe assembly

A silicon wafer was first patterned with 1 μm thick areas of gold via photolithography combined with electron beam evaporation, electroplating and wet etching. A few μg of (nominally) 100 nm diameter multi walled CNTs (Sigma-Aldrich) were dispersed in propan-2-ol by a horn ultrasonicator (Branson SLPt) and then pipetted onto the heated (60 $^oC$) pre-patterned silicon wafer. The heating ensured that the propan-2-ol evaporated quickly and results in uniform CNTs distribution over the wafer, that was then chemically etched in a $XeF_2$ etcher (Xactix Inc) to form suspended regions of gold with protruding CNTs (Fig. 1a).

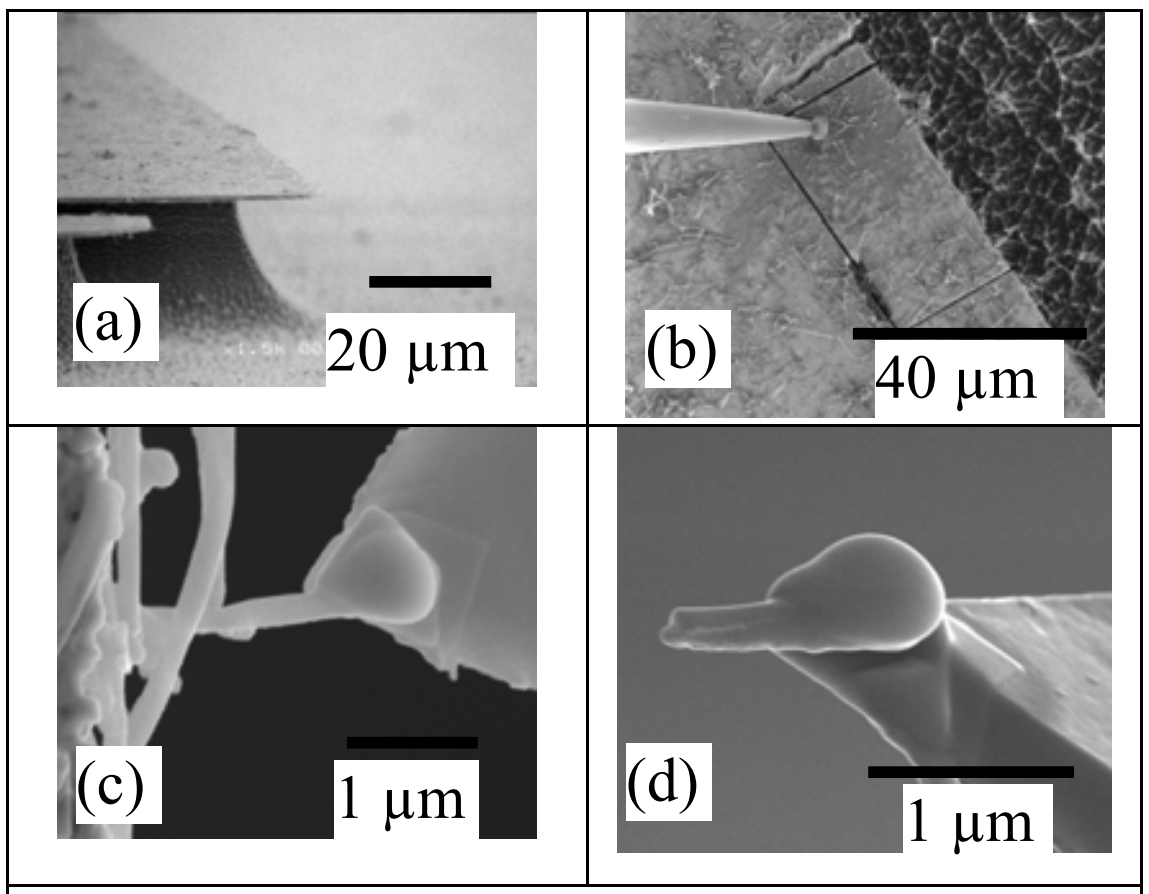

**Fig. 1.** CNT-SThM probe assembly showing: (a) CNT deposition onto the Au coated Si stub; (b) removal of small square of that stub with CNTs sticking out over the edge for attachment using a nano-manipulator; (c) attachment of CNT by Pt deposition to the SThM tip; and (d) ion beam cutting of CNT.

The tip of a nano-manipulator (Omniprobe, Oxford Instruments) mounted within a focused ion beam (FIB) system (Helios Nanolab 600 dual beam, FEI) was welded to a suspended gold region via ion-beam induced Pt deposition. The FIB was then used to cut this section of gold free from the substrate (Fig. 1b) allowing the attached foil to be moved by the nano-manipulator. A single CNT that protruded from the edge of the Au foil was then chosen, brought into contact with the end of SP type SThM probe (Kelvin Nanotechnologies), and subsequently welded to its apex by e-beam induced Pt deposition (Fig. 1c). The CNT was then cut by FIB at an angle, ensuring smaller CNT sample contact area, and the Au foil was pulled away, leaving the CNT attached protruding from the SP SThM probe end (Fig. 1d). The whole assembled CNT-SThM probe is given in Fig. 2, where the insets show the probe apex before and after contact AFM and SThM imaging. These indicate no observable changes in the probe geometry and confirm robustness of the probe.

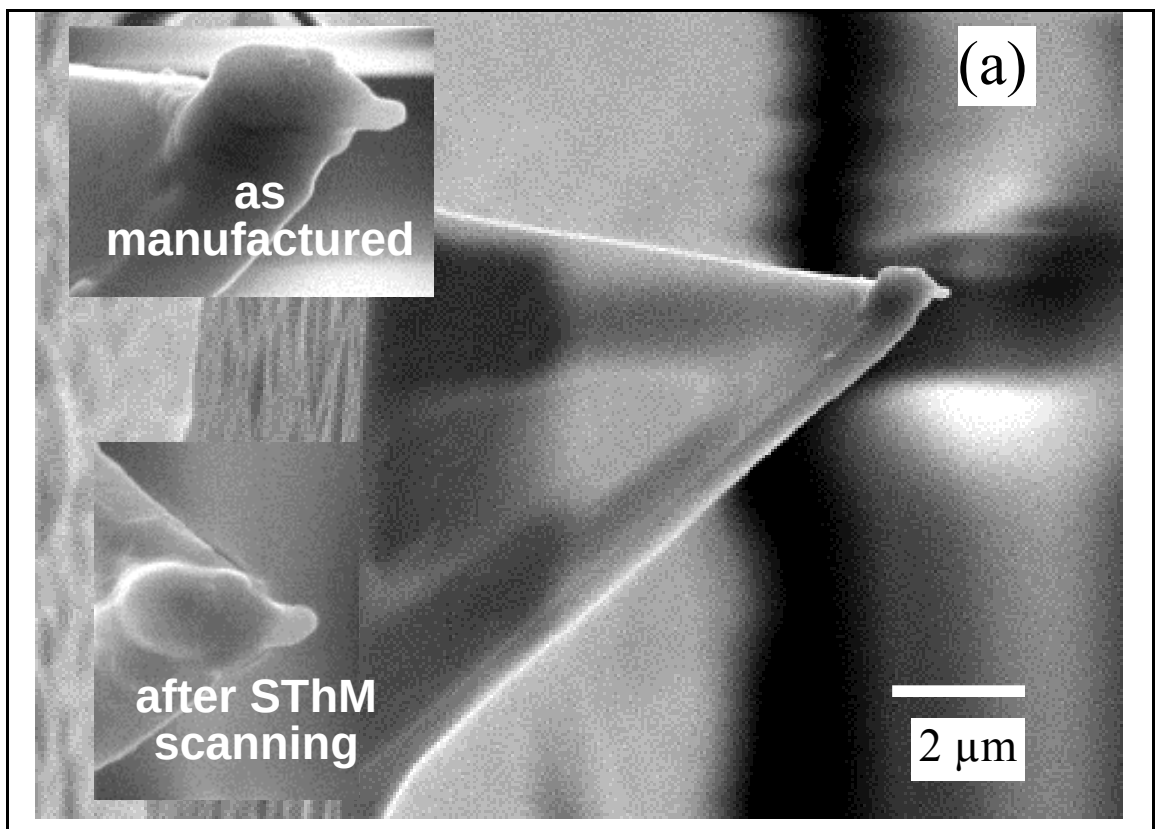

**Figure 2.** SEM images of the assembled of CNT-SThM probe (a). Insets show the same probe before after contact SThM scanning, indicating no observable damage to the probe.

**2. SThM setup, thermal calibration and measurements**

The CNT-SThM probes resistance-*vs*-temperature response was calibrated at 7 temperature points between 20 °C to 80 °C by thermal contact of the whole sensor with a Peltier hot/cold plate (Torrey Pines Scientific, EchoTherm-IC20) measuring DC resistance using 4 probe multimeter (Keithley 2100) and 90 kHz AC resistance in Maxwell bridge configuration (Fig. 3). In AC mode function generator (Keithley 3390) was used for bridge excitation and lock-in amplifier (LIA) (Stanford Research Systems, SRS-830) for detection of bridge output. During calibration two regimes were used –first, a low voltage regime eliminated probe self-heating linked probe resistance with the temperature, then an increased DC offset in both modes was used resulting in the self-heating, the self-heating temperature was obtained via first stage low-voltage calibrations. All these calibrations took place outside the AFM so there was no additional heating from the laser. For real-time scanning SThM measurements, AC approach was used, with the bridge balanced before each measurement, providing the real-time temperature reading of the SThM probe.

Both AFM and SThM imaging were performed in the standard SPM setup (Bruker Multimode, 100 µm scanner, Nanoscope IIIa controller, signal access module for readout of external signals) with SThM probe adapter (Anasys Instruments). The measurement bridge was balanced before each measurement to compensate for the ambient temperature drift and the laser heating of the probe, with LIA output acquired by the SPM during the scanning. The applied DC offset was kept constant during the imaging, resulting in practically constant power dissipated by the probe. In this configuration, the increased heat transport to the sample (due to local higher thermal conductivity) results in lower probe temperature, and darker areas in SThM images.

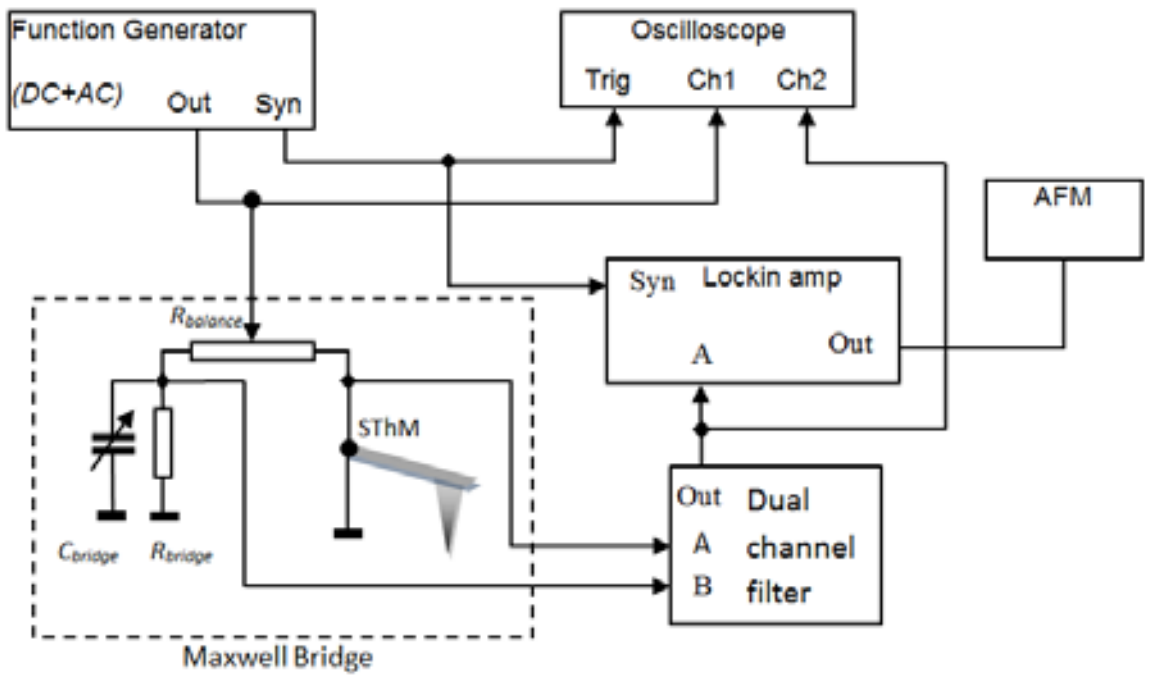

**Fig. 3**. Schematic diagram of SThM measurement electronics including the Maxwell bridge, the generator applying AC voltage and DC offset; bridge AC output detected by the lock-in amplifier is a linear function of the thermal sensor temperature.

## 3. Modeling of CNT-SThM probe

### 3.1 Finite element analysis simulations

The COMSOL Multiphysics finite element (FE) simulations were created in 3D with an air block encasing the whole probe. The dimensions of the air block and mesh step was changed by the factor of 2 and 4 resulting in the minimal temperature difference change by 0.3 to 1 % confirming that used settings were appropriate. The probe apex tip was either a CNT contacting the $Si_3N_4$ SThM probe along 1 um length or a probe itself. All outer boundaries were thermally anchored to room temperature of 293 K[26]. The thermal contact resistance of the CNT to the SThM probe was shown to have little effect on temperature for low and high thermally conducting samples due to the large surface area of the side of the CNT in contact with the tip. At the same time the contact resistance between the CNT tip and samples had a larger effect, adding a scaling factor of 0.7-0.9 for low thermal conductivity materials (~1 $Wm^{-1}K^{-1}$) like polymers, changing to scaling factor of 0.5 for high thermally conducting samples like Si (~100 $Wm^{-1}K^{-1}$). Such reduction, while notable, still preserves sufficient thermal signal and is fully counteracted by the decrease of the thermal resistance of the contact resistance (see below analysis in II.3.2) if CNT is used.

### 3.2 Analytical model of the CNT-SThM probe

In order to compare the contribution of different thermal resistances in the CNT-SThM probe assembly, *i.e. the* thermal resistances: from the heater to the cantilever base and air ($R_c$), from heater-to-CNT - $R_{CNT\text{-}heater}$ , CNT itself $R_{CNT,}$ and the thermal resistance of the sample $R_s$, we have built a simple analytical model of the CNT-SThM (Fig. 4). The heat flow from the heater passes through the thickness of the cantilever and bonding compound to reach the CNT normally with regard the nanotube axis, allowing to estimate the thermal resistance as[36]

$$R_{\text{CNT-heater}} = \ln\left(t_c / r_{\text{CNT}}\right) / \left(2\pi k_c L_{\text{CNT-heater}}\right) \quad (1)$$

where $k_c$ is the cantilever and bonding material thermal conductivity approximated as $Si_3N_4$ thermal conductivity ($k_{Si3N4} \approx 4$ $Wm^{-1}K^{-1}$), $t_c$ is cantilever thickness, $r_{\text{CNT}}$ – CNT radius, and $L_{\text{CNT-heater}}$ – length of the CNT-heater bond along CNT direction. The heat flow in the protruding CNT part is predominantly along CNT axis allowing this thermal resistance to be estimated as

$$R_{CNT} = k_{CNT} A_{CNT} / L_{CNT} \quad (2)$$

where $k_{CNT}$ is the axial thermal conductivity of the CNT, $L_n$ the CNT length and $A_{CNT}$ - the nanotube cross section area. By substituting the parameters with values for our probe (heater diameter and length ~500 nm, CNT diameter 100 nm and CNT thermal conductivity 1000 $Wm^{-1}K^{-1}$), estimates of these resistances are: $R_{CNT\text{-}heater} = 9.4\times10^4$ $KW^{-1}$, $R_{CNT} = 3.2\times10^3$ $KW^{-1}$. The thermal resistance of the CNT-sample $R_s$ has two parts: resistance due to the constriction of the narrow contact and the boundary resistance. The thermal resistance due to the tip-surface contact with effective radius $r_{cont}$ can be estimated as[16]

$$R_s = 2/3\pi k^* r_{cont} \tag{3}$$

where $k^*$ is the effective thermal conductivity of the contact without accounting for boundary resistance, defined as

$$k^* = 1/\left(k_s^{-1} + k_{tip}^{-1}\right) \tag{4}$$

where $k_s$ is the sample and $k_{tip}$ is the tip thermal conductivity. In the case of CNT-SThM probe and Si sample ($k_{si}$= 130 $Wm^{-1}K^{-1}$) thermal conductivity of Si defines the contact resistance $R_s$ (as $k_{tip}^{-1}$ is much smaller than $k_s^{-1}$) whereas for standard $Si_3N_4$ probe, the term of $k_{tip}^{-1}$ is dominating, masking the properties of the higher thermal conductivity materials. For Si sample and CNT tip a typical boundary thermal contact resistance [26] is $R_s$ = 3.0 x$10^6$ $KW^{-1}$. It should be noted that a parallel heat channel to the cantilever base and air ($R_c$) is approximately an order of magnitude lower with $R_c \approx 1.0\text{x}10^5$ $KW^{-1}$ (using FE calculated and experimentally measured values[26]) than the sample resistance, resulting in general reduction of the thermal response for any type of active probe and making any decrease of the probe-sample resistance (*e.g.* using the CNT probe) significantly improving performance of the SThM thermal measurements.

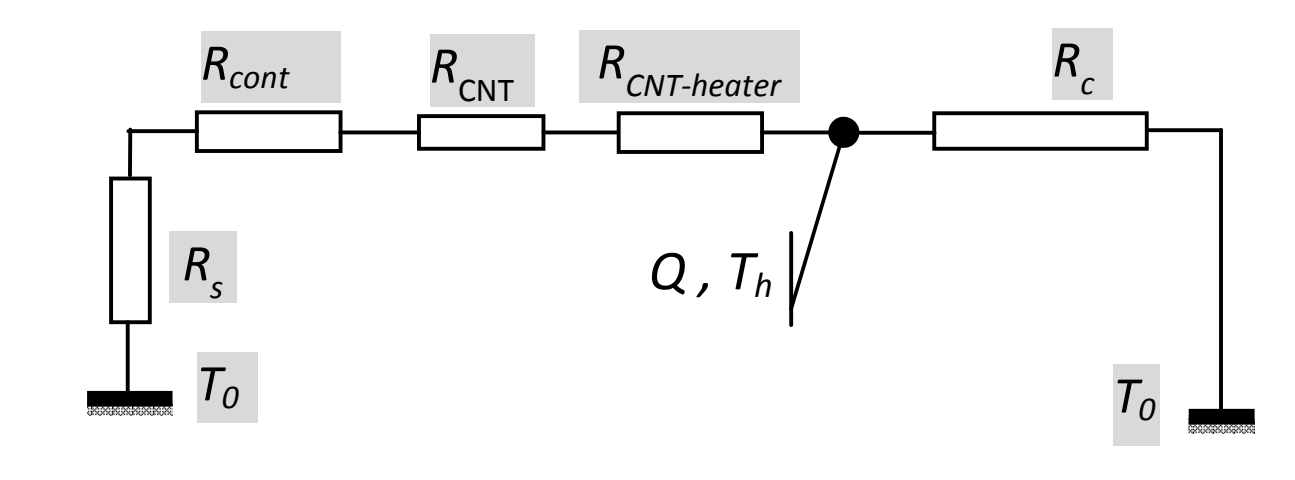

**Figure 4.** Analytical model of the SThM probe. Heat generated in the probe is dissipated tp the cantilever base and the air via thermal resistance

$R_c$, as well as to the sample via series of thermal resistances CNT-heater - $R_{CNT\text{-}heater}$, the resistance of the CNT tip itself $R_{CNT}$, the contact resistance $R_{CNT}$ and the resistance of the sample $R_s$. SThM measures the resulting temperature of the heater $T_\mathrm{h}$ that is function of the sample resistance $R_s$.

**4. Sample preparation ultra large scale integration and few layer graphene samples preparation**

In this study, we used test structures of ultra large scale integration (ULSI) interconnects (Fig. 5) containing Al leads embedded in a low-k polymer (benzo-cyclo-butene, or BCB) interconnects and 300 nm $SiO_2$ on Si wafer substrates for graphene deposition. All were sequentially sonicated in methanol, isopropanol, and DI water (10 min each), and cleaned in $Ar/O_2$ plasma cleaner (Zepto LF). The graphene sample was mechanically exfoliated onto the surface by using pressure sensitive tape immediately after plasma cleaning[19].

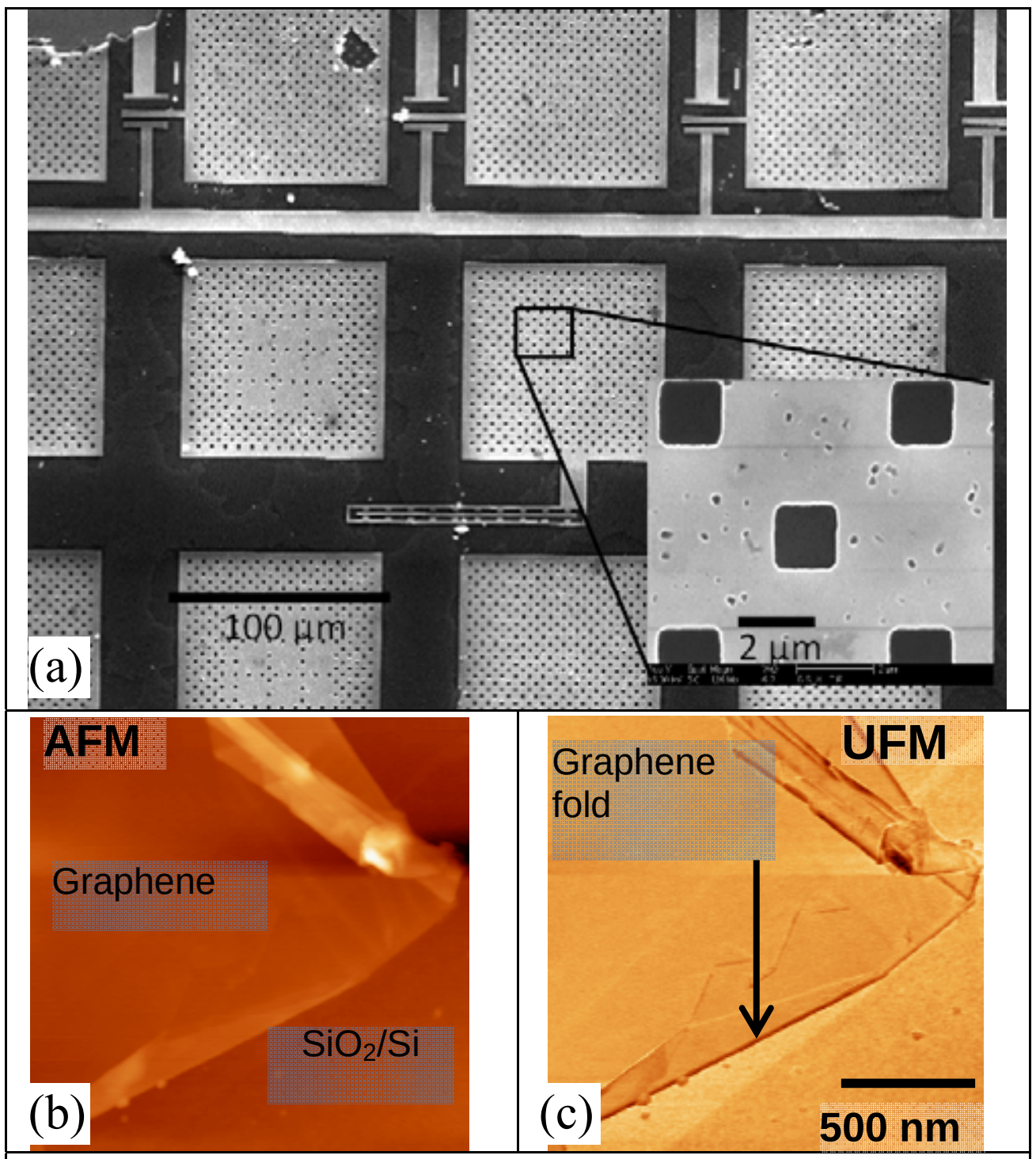

**Fig. 5.** a) SEM image of the test sample of chemo-mechanically planarization processed ULSI Al – metal leads embedded in the benso-cyclo-butene low-k dilectric; inset: higher magnification image showing polishing pits in the metal layer. b) AFM topography and c) ultrasonic force microscopy (UFM) nanomechanical properties maps of few layer graphene on Si showing graphene layer edge and folds; same area was imaged in SThM and CNT-SThM in Fig. 7

## III. Experimental results and discussions

As discussed above, CNTs should allow superior spatial resolution and thermal sensitivity for existing SThM probes, up to the limits defined by fundamental constraints such as the mean-free-path of the heat carriers (electrons in metals and phonons in insulators and semiconductors)[37, 38] and Kapitza contact resistance.[39] Nanotubes will also provide a high thermal conductance path between the µm sized thermal sensor and the sample, with highly thermally conductive CNT apex securing a low thermal contact resistance, allowing materials up to 100's W $K^{-1}m^{-1}$ to be studied as foreseen in the analytical studies elsewhere.[26] Furthermore, high mechanical strength of CNT's (Young's modulus in the TPa range[40, 41]) provides robust and hardwearing AFM tips.[42, 43] The CNT-SThM probe tip (Figs. 1, 2) is quite short and stumpy, but had good thermal anchoring and low thermal resistance to the tip due to the Pt metal welding and the long contact area. In addition, the small contact radius due to the sharpened apex provides an appropriate rigid tip for scanning.

## 1. Thermal mapping of ULSI interconnects

CNT-SThM was first used to study thermal transport in an ULSI interconnects made of Al metallization tracks embedded in the low-k dielectric BCB cross-linked polymer on a Si wafer. The entire structure was chemo-mechanically polished during the planarization process.[34] The thermal properties of this nanostructure are particularly important since they model real interconnects, where metal tracks carry high currents and their overheating increases electromigration and adversely affects the device performance. In this structure, both high and low thermal conductive materials are presented, side-by-side, producing clear boundaries and allowing the estimation of the sensitivity and spatial resolution of the CNT-SThM. Similar areas of ULSI interconnects were imaged using a standard SP SThM probe (Fig. 6a,b) and a CNT-SThM probe (Fig. 6c,d), respectively. The $Ar/O_2$ plasma mix that was used to clean the sample before scanning could have possibly etched the BCB slightly leading to a marginally increased height difference between the metal and BCB. This height difference creates a change in contact area when scanned over the metallization border giving the dark edge to the metallization track seen in both thermal images.

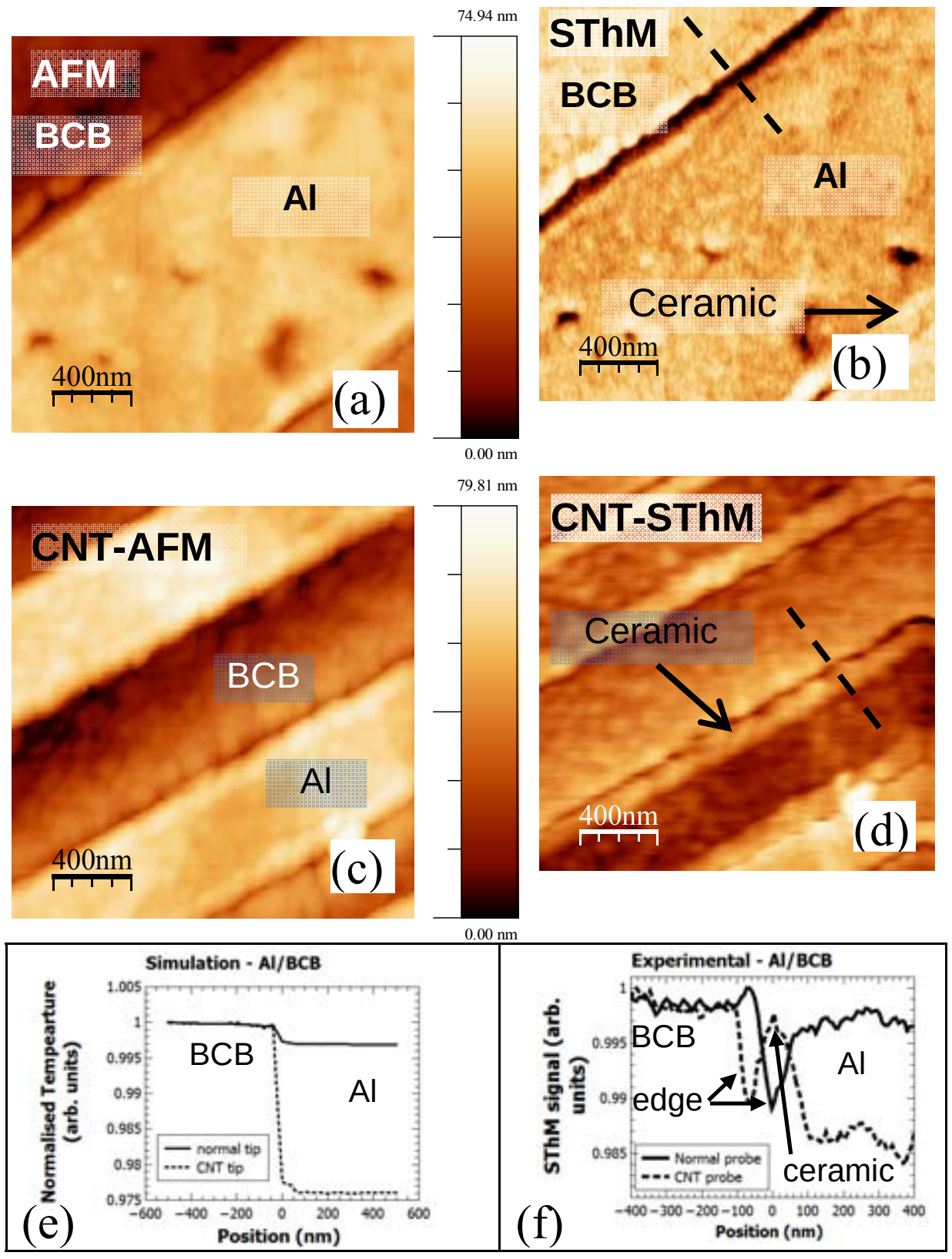

**Figure 6**. SThM images of Al-BCB ULSI interconnects sample in air. (a) Height, and b) – thermal image using a standard SP SThM probe; and (c) - height and (d) - thermal image using a CNT-SThM probe. With the constant power applied to the probe, the higher local thermal conductivity results in lower probe temperature, and correspondingly darker SThM contrast. The larger thermal conductivity of Al tracks is visible in both images, lower thermal conductivity in modified ceramic interlayer between metallization and BCB can be clearly observed only with CNT-SThM. (e), (f) Comparison of FE simulated thermal 1D profiles of the Al/BCB boundary with CNT-SThM probe and a standard probe (e) with experimentally obtained profiles (f). Dotting line mark where the profiles were taken.

The AFM and SThM data above show that the CNT probe exhibits a spatial resolution at least comparable with the standard probe revealing both polymer and ceramic layer fine structure, which are clearly visible on both the thermal and height images. Analysis of Fig. 6 indicates a ~20-30 nm spatial resolution in topography while the thermal resolution is of the order of 30 nm, on a par or better than achieved so far in SThM in air with the similar probes.[44] More significantly, while the larger thermal conductivity of metal tracks is clear in both images, the lower thermal conductivity in the modified ceramic interlayer between Al and BCB can be clearly observed only with the CNT-SThM. The CNT-SThM probe high spatial resolution is also seen in other AFM modes like friction AFM imaging modes that revealed the fine structure of the BCB polymer and the metallization tracks, including voids.

**2. Analysis of thermal contrast in CNT-SThM.**

Using FE modeling, we simulated CNT-SThM and a regular SP SThM probe one-dimensional scans across a polymer-metal interconnect boundary. Fig. 6 compares simulated and experimental results recorded under the similar conditions. The simulation clearly indicates that the theoretical sensitivity of the CNT-SThM probe is an order of magnitude better compared to that of the standard probe, in line with the theoretical analysis presented elsewhere[26] and in very good correlation between the simulation and experimental (Fig. 6 e, f) with the experimental data showing the ceramic interlayer at the interface between the BCB and Al not present in the simulation. The FE calculations did not take into account the interfacial Kapitza resistance, the influence of ballistic conductance or modification of the contact area due to the surface topography, all of which would contribute to the experimental features of CNT-SThM images. Nevertheless, the experimental results clearly confirm that CNT-SThM outperforms the standard SThM probe.

Given the typical values for the CNT-SThM probe components (CNT diameter 100 nm and thermal conductivity 1000 $Wm^{-1}K^{-1}$), the thermal resistances were estimated as; $R_{CNT\text{-}heater}$ (heat flow from the heater to the CNT) - $9.4x10^4$ $KW^{-1}$; $R_{CNT}$ (heat flow along the CNT) - $3.2x10^3$ $KW^{-1}$; and $R_s$ (tip-sample contact) - $3.0x10^6$ $KW^{-1}$ as described in Methods section II.3.2. The high thermal conductivity of the CNT probe makes the thermal conductivity of the contact practically independent of the probe conductivity even for such high thermal conductivity materials as Si, GaAs and Al, which offers additional advantages for metrological applications. For very small nanoscale contacts and for highly crystalline materials, the considerations above will be affected by the mean–free-path (MFP) of the acoustic phonons $\Lambda$, when it becomes larger than the contact radius $r_{\text{cont}}$. Theoretical analysis predicts a different thermal resistance[45] depending on a Knudsen number, $\text{Kn} = \Lambda / r_{cont}$ at Kn>>1. In this case, the effective contact thermal conductivity in the intermediate diffusive-ballistic regime $k^{*}_{cont\text{-db}}$ can be estimated as the decrease of the effective heat conductivity of the contact $k^{*}_{\text{cont}}$,

$$k^{*}_{cont-db} = \frac{k^{*}_{cont}}{2(\pi Kn)^2}\left[\sqrt{1+(2\pi \text{Kn})^2}-1\right] \tag{5}$$

Low Knudsen numbers (Kn << 1) correspond to diffusive transport whereas Kn >> 1, describe the ballistic regime. For a $Si_3N_4$ tip, Kn ≈ 0.3-0.5, giving diffusive regime. In turn, the CNT-SThM tip is firmly in the ballistic regime by virtue of the CNT phonons MFP often in the range of 250-700 nm,[46, 47] and the associated Kn ≈ 5-10. While heat transport as the energy loss from the probe to the sample is objectively measured in any regime, the ballistic nature of heat transport at the distances of few tens of nm may pose a fundamental limitation to quantitative measurements of local thermal conductivity.

## 3. CNT-SThM mapping of few layer graphene

As discussed, one of key advantages of the CNT-SThM probe is its ability to distinguish between high thermal conductivity materials. Since the in-plane thermal conductivity of graphite and few layer graphene (FLG)[19, 48, 49] have one of the highest known in nature thermal conductivities (on the order of 2000-5000 $Wm^{-1}K^{-1}$), these materials were chosen to investigate the performance of the CNT-SThM probe. In addition, a sharp edge of a graphene flake can provide a good feature for determining the spatial resolution. Fig. 7 clearly shows that the CNT-SThM of a 3 nm thick graphene flake on 300 nm $SiO_2$ on a Si wafer substrate has a much better resolution both in AFM and SThM modes. Unlike the standard SThM micrographs (Fig. 7a, b), the CNT-SThM probe recorded the layered steps in the graphene along the right hand edge, with a much better lateral resolution (Figure 7c, d). Likewise, the standard SThM probe was less thermally sensitive to the graphene layer, with a lower contrast between the graphene and the substrate (Fig. 7b). Higher thermal contrast images are obtained for the CNT-SThM probe, with a clear edge and a strong contrast between graphene and substrate. The simulation data show even the more pronounced trend as thicker graphene layer had to be used for FE simulation to allow convergence of the model. In addition to the lower thermal resistance of the CNT tip, the lower phonon mismatch between graphene and CNT that defines the interfacial Kapitza resistance,[50, 51] may have played a significant role in improving the contrast. Finally, the contact between the CNT tips and graphene leads to a much less spurious height influence.

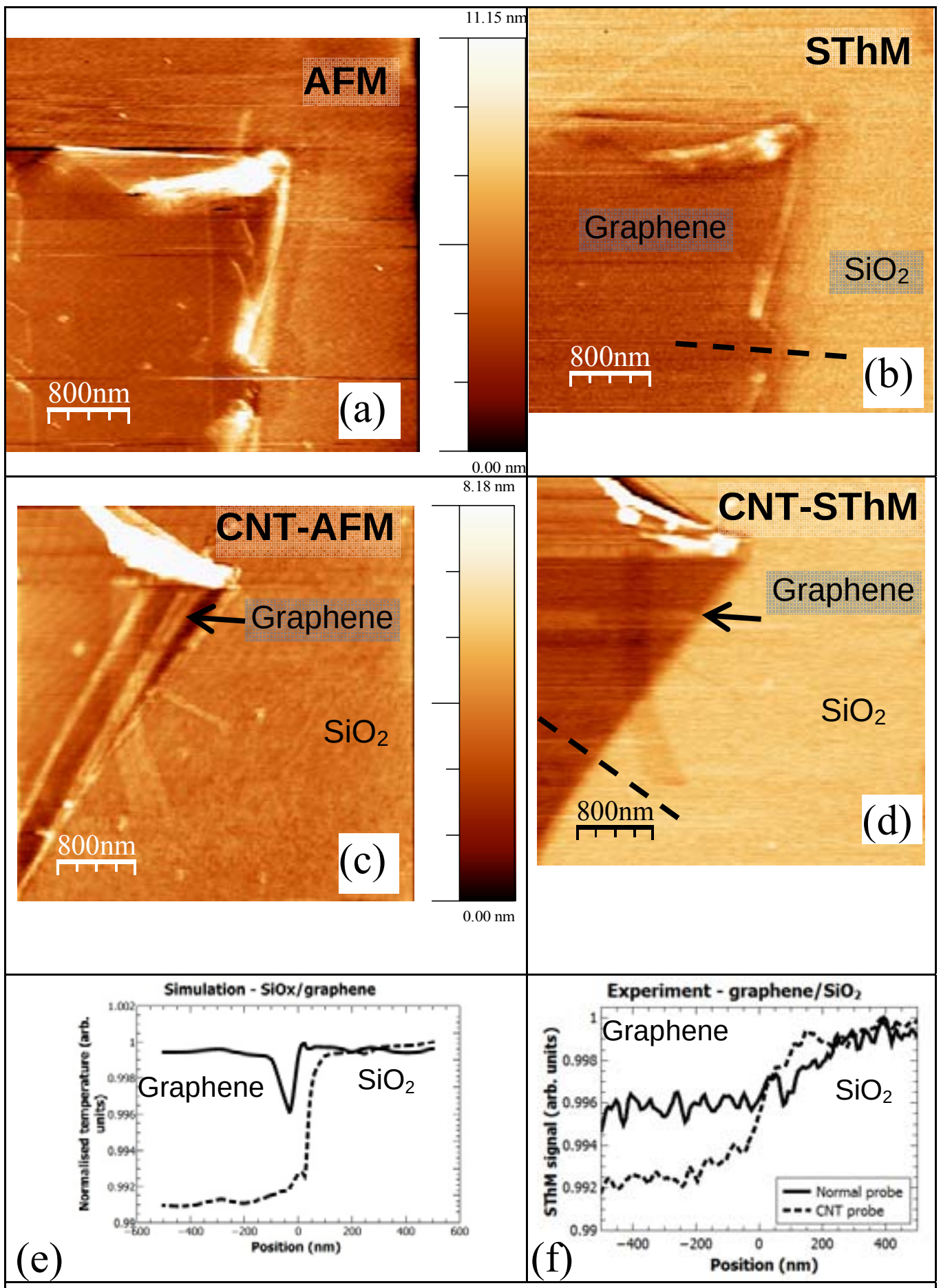

**Figure 7.** SThM imaging of few layer graphene flake of 3 nm thickness on Si substrate using a standard SThM probe with height (a) and thermal image (b), compared to same area scan using a CNT-SThM probe topographic height (c) and thermal signal (d). Although the same flake was imaged, some minor displacements may have occurred between the scans (a) and (c). (e) One-dimensional sections (profiles) of the flake edge show much higher spatial resolution of thermal signal in CNT-SThM with FWHM of 30 nm as well as less noisy thermal signal. The dotted line marks where the profiles were taken.

The reference AFM tapping mode and ultrasonic force imaging[52] of the graphene steps of sample shown in Fig. 5 verified their topography. As mentioned above, pre and post-scanning SEM images (insets in Fig. 5) of the CNT-SThM tip showed no changes to the tip shape confirming the robustness of CNT-SThM probe.

## 4. Future prospects of CNT-SThM

Due to the CNT's highly anisotropic thermal conductivity, it acts as a guide for the heat travelling along its length. This would make SThM imaging in, for example a liquid environment, possible with the heat delivered into the tip-sample junction while reducing heat losses to the environment. At the same time, some CNT-SThM limitations need to be considered. The first of these is the thermal contact resistance between the tip and the surface which can modify the probe response. Finite element models conducted with the contact resistance shows that for low conductivity samples, the effect was small. However, for highly thermally conducting samples, *e.g.* Si, the sensitivity could be reduced by up to 50%. Furthermore, in an arbitrary sample, the tip-surface thermal resistance depends strongly on the contact area nanoscale geometry, that can be modified during scanning due to topographical variations of the sample, and can also be affected by the anisotropy of thermal conductivity. Likewise, the large ballistic phonon MFP, inherent to the ballistic nature of thermal transport in CNT probes may constitute a further limit due to phenomena such as interface scattering. While all these aspects are to be considered, especially for the use of CNT-SThM for quantitative measurements of heat transport, the results presented here demonstrate clearly that CNT-SThM probes is a way forward for investigating high thermal conductivity materials such as those used in semiconductor industry and nanoscale devices and offers a great improvement over standard SThM probes.

# IV. Conclusions

We have demonstrated a novel CNT-SThM probe using a carbon nanotube attached to the tip of a conventional SThM probe. The mapping of thermal transport in ULSI interconnects revealed a fine structure of ceramic layers between metallization and dielectric that may provide a heat barrier and increase electromigration in such nanostructures. For 3 nm thin graphene flakes, the CNT-SThM sensitivity to heat transport and the spatial resolution of a new probe, is superior to the standard probes.

The new thermal probe can sense both low thermally conducting polymers and highly thermally conducting metallization and graphene, with contrast exceeding that of standard SThM probes, and reduced topographical artifacts. A thermal resolution of 30 nm was achieved for the ULSI structures with similar spatial resolution in topographic and friction images. SEM inspection of probes before and after imaging confirmed that CNT-SThM probes are hard wearing and robust. CNT's extreme axial thermal conductivity that links the thermal sensor and the nanoscale sizes area of the sample providing high sensitivity to the sample thermophysical properties, whereas its high aspect ratio and robustness can make it the perfect thermal probe for the future generation of SThM.

## Acknowledgments

Authors acknowledge input of Craig Prater and Roshan Shetty from Anasys Instruments for the support related to the SThM development. We acknowledge the support from the EPSRC grants EP/G015570/1 and EP/G017301/1, EPSRC-NSF grant EP/G06556X/1 and EU FP7 grants, GRENADA (GA-246073) and FUNPROB (GA-269169). We would also like to thank the group that produced WSxM for their useful program for analysis of SPM images.[53]

**The TOC entry**

A scanning thermal microscope with a carbon nanotube tip was used to map thermal conductivity in ultra-large scale integration interconnects and a few layer graphene with 30 nm spatial resolution.

Peter D. Tovee, Manuel E. Pumarol, Mark C. Rosamond, Robert Jones, Mike C. Petty, Dagou A. Zeze and Oleg V. Kolosov

Ultra High Thermal Resolution Scanning Probe Microscopy via Carbon Nanotube Tipped Thermal Probes

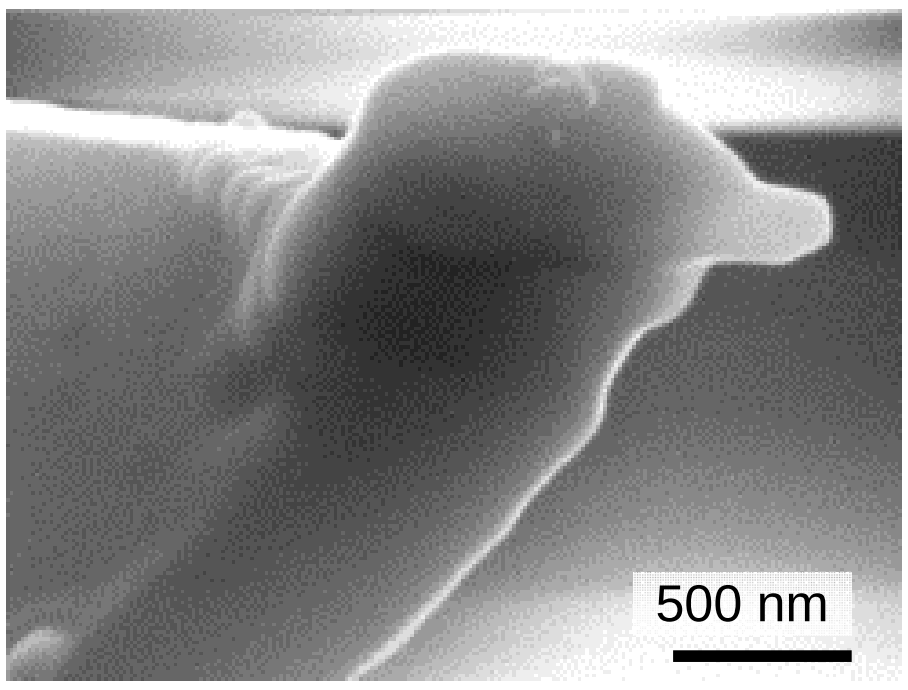

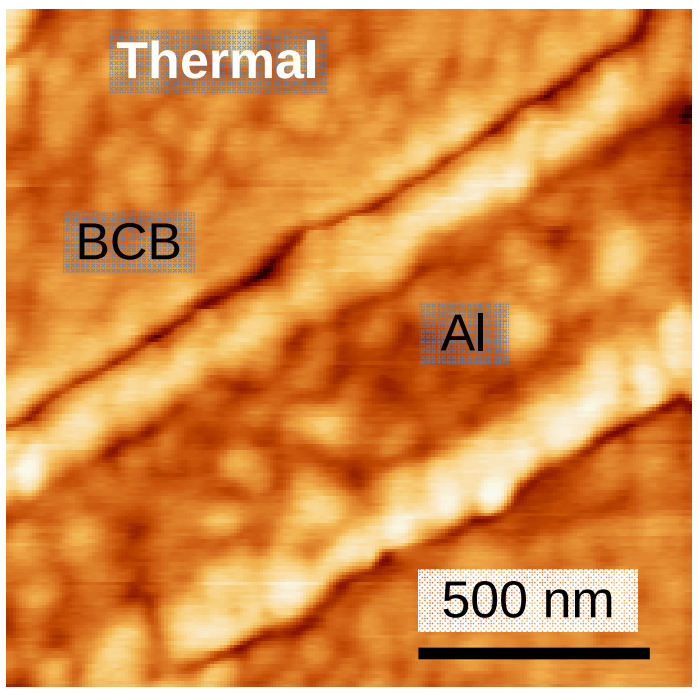